\documentclass[aps,prb,twocolumn,showpacs,superscriptaddress,groupedaddress]{revtex4-2}
\usepackage{graphicx,color}
\usepackage[english]{babel}
\usepackage{hyperref}
\definecolor{link}{rgb}{0.1,0.1,0.9}
\hypersetup{colorlinks=true,linkcolor=link,citecolor=link,urlcolor=link,linktocpage}
\usepackage{epstopdf}
\usepackage{amsmath}
\usepackage{amssymb}
\usepackage{longtable}
\usepackage{mathtools}
\usepackage{float}
\usepackage{subfigure}
\usepackage{booktabs}
\usepackage{adjustbox}
\usepackage{multirow}
\usepackage{dcolumn}
\usepackage{tabularx}
\usepackage[normalem]{ulem}

\setcitestyle{numbers,square}

\setcitestyle{numbers,square}

\begin{document}


\title{Anisotropy driven critical phenomena in a van der Waals ferromagnet}

\author{Neesha Yadav}
 \altaffiliation[Corresponding author:]{nishuyadav0311@gmail.com}
\affiliation{Department of Physics, Indian Institute of Technology Delhi, Hauz Khas, New Delhi-110016, India
}
\author{Harish Chandr Chauhan}
\altaffiliation[Corresponding author:]{harishcc02901@gmail.com}
\affiliation{Department of Physics, Indian Institute of Technology Delhi, Hauz Khas, New Delhi-110016, India
}

\author{Pintu Das}
 \altaffiliation[Corresponding author:]{pintu@physics.iitd.ac.in}
\affiliation{Department of Physics, Indian Institute of Technology Delhi, Hauz Khas, New Delhi-110016, India
}

\date{\today}

\begin{abstract}
We report a detailed investigation into the field-dependent critical behavior of the van der Waals ferromagnet $\mathrm{Fe_3GeTe_2}$ to elucidate how applied magnetic fields ($B$), strong magnetocrystalline anisotropy, and emerging other energy scales modulate its critical behavior. Utilizing the recently developed method, viz., the average of normalized slope (ANS), for analysis of critical behavior, we evaluate the critical exponents ($\beta$, $\gamma$, and $\delta$) across distinct field windows up to $\rm{7\,T}$. For $B$ applied along the crystal's easy axis ($B\parallel c$), the system exhibits long-range interactions dominated by competing double-exchange and itinerant Ruderman-Kittel-Kasuya-Yosida (RKKY) mechanisms, which is systematically suppressed with increasing $B$ to converge toward the long-range 3D Ising universality class ($\beta\sim0.330$, $\gamma\sim1.00$). Conversely, under an in-plane hard-axis configuration ($B\perp c$), the critical behavior study exhibits a sub-unity regime ($\gamma<1$) across the measured range of $B$. This behavior is attributed to the intense competitive interplay between the external Zeeman energy and strong out-of-plane magnetocrystalline anisotropy. Our findings demonstrate that critical behavior in nontrivial van der Waals
magnets is fundamentally modulated by the orientation of $B$ and the strength of magnetocrystalline anisotropy, demonstrating the necessity of $B$-dependent critical behavior investigation over conventional frameworks.
\end{abstract}
\maketitle

\section{Introduction}

The investigation of critical behavior in low-dimensional van der Waals (vdW) materials provides vital insights into the mechanisms driving the magnetic ordering, the spatial range of exchange coupling, and the underlying magnetic universality class (UC) of a system~\cite{stanley1999scaling, fisher1974renormalization}. A UC formally defines the space-dimensionality ($d$) and spin-dimensionality ($n$) of the local magnetic moments. In the vicinity of a continuous paramagnetic phase transition ($T_{\rm{C}}$), the thermodynamic response functions$-$such as susceptibility ($\chi$) and specific heat$-$exhibits power-law scaling governed by a set of critical exponents (CEs)~\cite{stanley1999scaling, fisher1974renormalization, phan2010tricritical, tiwari2020ising, scheie2022spin, parhizgar2013indirect, geldenhuys1978rkky, bisht2023investigation, chauhan2026magnetic, Liu2017, liu2017critical}. These exponents are uniquely determined by the symmetry of the order parameter, the dimensionality of the system, and the range of the underlying exchange interaction, classifying the phase transition into standard UCs such as the Ising ($n=1$), XY ($n=2$), Heisenberg ($n=3$), or mean-field models~\cite{stanley1999scaling, fisher1974renormalization, chauhan2026magnetic, Liu2017, liu2017critical}. Ideally, the Heisenberg UC describes the critical behavior of isotropic magnetic systems, the XY UC explains that of magnetic systems with planar anisotropy, and the Ising UC applies to that of magnetic systems with uniaxial anisotropy~\cite{tiwari2020ising}.

Conventionally, these CE values are extracted from the magnetic field ($B$)-dependent magnetization ($M$-$H$) isotherms near $T_{\rm{C}}$ using the conventional modified Arrott plot (MAP) analysis or the Kouvel-Fisher (KF) method~\cite{ANPhysRevLett.19.786, Kouvel1964, Zhu2023, Gupta2018, Tateiwa2014, Bhattacharyya2011}. While these methodologies are strictly applicable to trivial magnetic systems, such as those governed by the Hamiltonian ($H=-J\sum_{i=1}^{n} \vec {S_i}.\vec S_j$), they frequently yield inconsistent results when applied to a nontrivial magnetic system~\cite{chauhan2026magnetic, yadav2026intrinsic, deng2024magnetic, zhang2013influence, wang2022competing, clements2017critical, tozri2011critical, dash2019impression, madhogaria2019evidence, heilman2000field, wang2023field}. In 2022, Chauhan \textit{et al.}~\cite{chauhan2022different} proposed a modified iteration method to address the erroneous CE values, obtained from the conventional critical analysis process, in nontrivial magnetic systems by estimating separate sets of CE values using $M$-$H$ isotherms for $T\leq T_{\rm{C}}$ and $T\geq T_{\rm{C}}$. Subsequently, Chauhan \textit{et al.}~\cite{chauhan2026magnetic} introduced intercept analysis of standard short-range (SR) and mean-field UCs alongside the average of normalized slope (ANS) method to evaluate the reliability and dominant exchange interactions of critical behavior analyses, validating these approaches on trivially treated Ni and nontrivial Gd.

The vdW ferromagnet $\mathrm{Fe_3GeTe_2}$ (FGT) represents a complex nontrivial magnetic system~\cite{yi2016competing, Deng2018, Tan2018,  Fei2018, liu2017critical, Liu2017,10.1063/1.4961592, AnkitaPhysRevB.109.L020407,Kim2026, yadav2026intrinsic, zhang2018emergence, sharma2023non}. It features a rich interplay of localized and itinerant electron magnetism, a strong out-of-plane (OOP) uniaxial magnetocrystalline anisotropy, and an intrinsic competition between interlayer ferromagnetic and low-temperature antiferromagnetic exchange channels. Consequently, prior efforts to investigate critical behavior have led to conflicting conclusions~\cite{liu2017critical, Liu2017, AnkitaPhysRevB.109.L020407}. An early study by Liu \textit{et al.}~\cite{liu2017critical} reported CE values ($\beta \sim\,0.327$, $\gamma \sim 1.079$, and $\delta \sim 4.261$) that did not belong to any standard UC, attributing the discrepancy to a three-dimensional (3D) long-range (LR) magnetic coupling that decays with distance $r$ as $J(r) \sim r^{-4.6}$. Notably, CE values do not necessarily have to belong to a standard UC when investigating the critical behavior of a non-trivial magnetic system~\cite{chauhan2026magnetic}. Investigations on $\rm{Fe}$-deficient variant ($\mathrm{Fe_{3-x}GeTe_2}$ with $\mathrm{x \approx 0.36}$)  yielded heavily altered CE values ($\beta\,\sim\,0.372$, $\gamma\,\sim\,1.265$, $\delta\,\sim\,4.50$), highlighting a strong sensitivity to local vacancies and showing 3D Heisenberg UC with weaker LR magnetic coupling that decays with distance $r$ as $J(r) \sim r^{-4.89}$~\cite{Liu2017}. More recently, reports by Ankita \textit{et al.}~\cite{AnkitaPhysRevB.109.L020407} argued for a two-dimensional Heisenberg exchange with LR interaction (Table \ref{Table_Critical exponents}) and discussed the violation of the Hohenberg-Mermin-Wagner theorem~\cite{PhysRevLett.17.1133, PhysRev.158.383}. N. D. Mermin and H. Wagner~\cite{PhysRevLett.17.1133} proved theoretically that it is impossible to stabilize long-range antiferromagnetism or ferromagnetism in one- or two-dimensional systems hosting isotropic Heisenberg models. Alternatively, this implies that long-range antiferromagnetism or ferromagnetism can exist in one- or two-dimensional systems hosting Ising or XY models. P. C. Hohenberg proved that LR order cannot exist in one- and two-dimensional Bose and Fermi systems at finite temperatures~\cite{PhysRev.158.383}. The experimental studies show that the CE values for FGT and its stoichiometric variants vary with $\mathrm{Fe}$ content, vacancies, and the chosen $B$ windows for investigations, thereby failing to demonstrate scaling convergence. To resolve these discrepancies and clarify whether these variations represent intrinsic physics or analytical artifacts, a $B$- and $B$-direction-dependent critical analysis is required.

In this article, we present a detailed investigation into the critical behavior of FGT using recently developed methodologies~\cite{chauhan2026magnetic}. BY analyzing both OOP ($B \parallel c$) and in-plane (IP) ($B \perp c$) magnetization isotherms, we quantify $B$-driven crossover between dominant LR and SR exchange interactions. Our results show that the CE values of FGT are $B$-dependent, shedding light on the fragile balance of competing energy contributions in anisotropic vdW magnets.

\section{Experimental details}
The chemical vapor transport (CVT)-grown single crystals of FGT were obtained from commercial sources, viz., HQ graphene (\url{https://www.hqgraphene.com/}). The magnetic measurements were carried out using a Magnetic Property Measurement System (MPMS3) (make: Quantum Design). Temperature-dependent magnetization ($M$-$T$) measurements were conducted under zero-field-cooled warming (ZFCW) and field-cooled warming (FCW) protocols. We recorded two sets of $M$-$H$ isotherms in the vicinity of $T_{\rm{C}}$ by applying $B \parallel c$ and $B \perp c$, where $c$ represents the FGT crystal's $c$-axis. 

\section{Results and Discussion}

\subsection{Magnetic properties of FGT}

FGT is predominantly a ferromagnet hosting two types of $\mathrm{Fe}$ ions, viz., $\mathrm{Fe^{3+}}$ and $\mathrm{Fe^{2+}}$. $\mathrm{Fe^{3+}}$ ($\mathrm{3d^5}$) is coupled with $\mathrm{Fe^{2+}}$ ($\mathrm{3d^6}$) via $\mathrm{Te}$ following double exchange mechanism. There is direct exchange coupling between $\mathrm{Fe^{3+}}$ ions, resulting in weak intralayer antiferromagnetism. The interlayer magnetic interaction is weak and can be easily modulated by applying a small $B$ (the Zeeman energy)~\cite{yadav2026intrinsic}. The AFM exchange breaks first around 160\,K, leading the system into a predominantly FM state near $T_{\mathrm{C}}$.

\begin{figure}[h!]
    \centering
    \includegraphics[width=\linewidth]{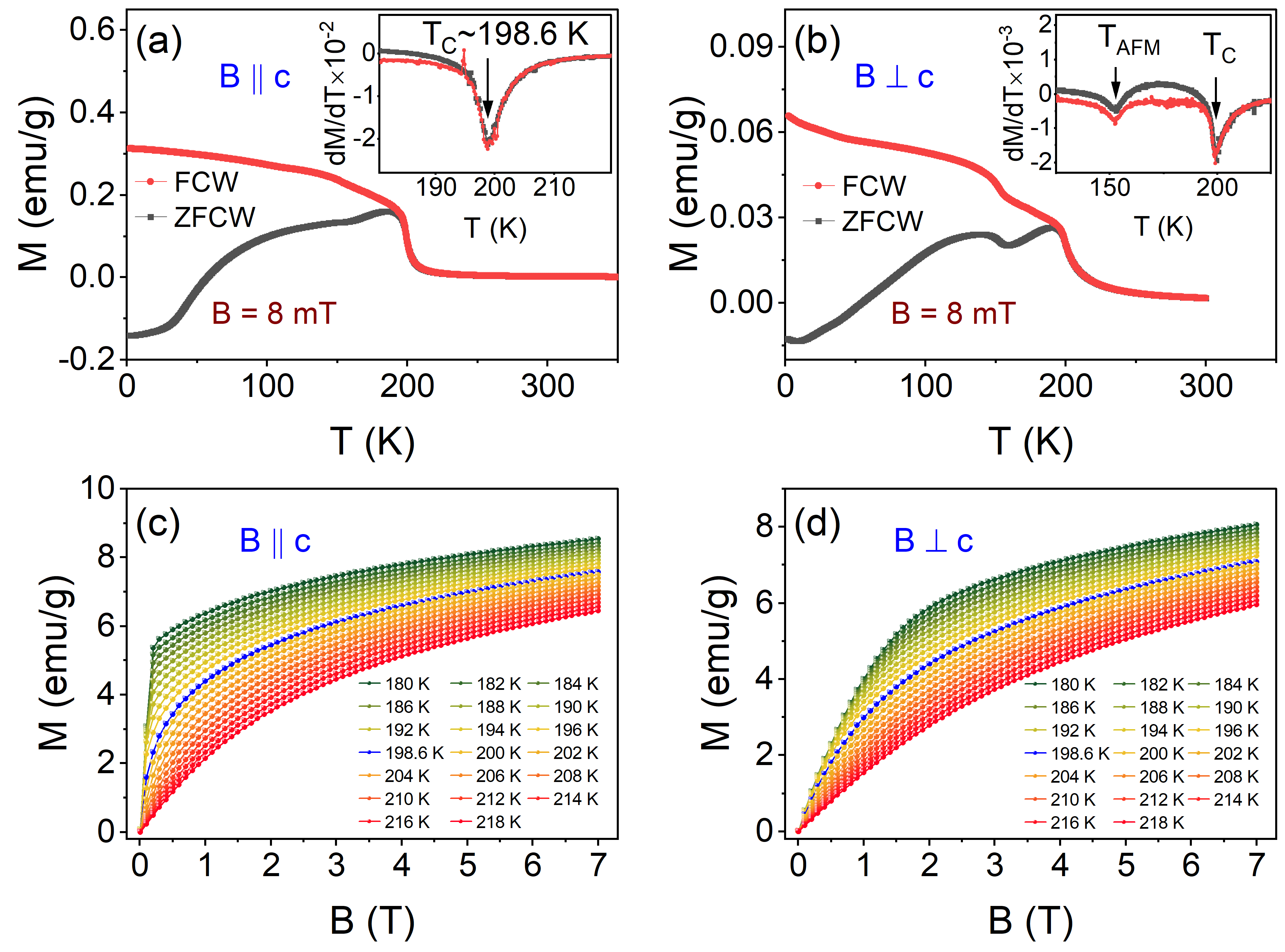}
    \caption{Temperature-dependent magnetization ($M$-$T$) recorded in zero-field-cooled (ZFC) and field-cooled (FC) warming mode by applying a magnetic field ($B$) of $\mathrm{8\, mT}$ (a) along the crystal's c-axis ($B \parallel \mathrm{c}$) and (b) perpendicular to the crystal's c-axis ($B \perp \mathrm{c}$). The inset of the respective figures shows the first derivative of the $M$-$T$, whose minima represent the Curie temperature, $T_{\mathrm{C}}=198.6\, K$. The $B$-dependent magnetization ($M$-$H$) isotherms recorded in the vicinity of $T_{\mathrm{C}}$ by applying (c) $B \parallel \mathrm{c}$ and (d) $B \perp \mathrm{c}$.}
    \label{MT_MH_FGT}
\end{figure}

We record the ZFC and FC $M$-$T$ [Figs. \ref{MT_MH_FGT}(a) and \ref{MT_MH_FGT}(b)] in the warming mode by applying $B$ of 8\,mT along both parallel ($B\parallel c$) and perpendicular ($B\perp c$) to the crystal's c-axis. The observed negative magnetization in the ZFCW curve results from dominant AFM exchange at low temperature and, to some extent, the experimental design, which may stabilize the magnetization in the negative direction via a partially induced diamagnetic interaction. The OOP ($B\parallel c$) and IP ($B\perp c$) $M-T$ plots confirm that FGT exhibits an easy axis along the crystal's c-axis, since the IP magnetic moment at 10\,K is $\sim50$ times lower than the observed OOP magnetic moment; see Figs.\,\ref{MT_MH_FGT}(a) and \ref{MT_MH_FGT}(b). When $B\perp c$ [Fig. \ref{MT_MH_FGT}(b)], a dip in the $M-T$ is clearly visible near $\mathrm{160\,K}$ which is the temperature where AFM coupling breaks and leads the system to a predominantly FM phase~\cite{yadav2026intrinsic, yi2016competing}. The minimum of the first derivative of $M$-$T$ [inset of Figs. \ref{MT_MH_FGT}(a) and \ref{MT_MH_FGT}(b)] yields $T_{\mathrm{C}}=\mathrm{198.6\,K}$ for both $B\parallel c$ and $B\perp c$. The observed lower $T_{\mathrm{C}}$ from the $T_{\mathrm{C}}$ value of intrinsic FGT suggests that the used crystal may have minor $\mathrm{Fe}$ vacancies~\cite{yadav2026intrinsic}. Therefore, the existing AFM and FM exchange interactions~\cite{yadav2026intrinsic}, the coexistence of itinerant and localized electrons~\cite{sharma2023non}, and emerging $\mathrm{Fe}$ vacancies make FGT a nontrivial magnetic system~\cite{chauhan2022different, chauhan2026magnetic}. We investigate how these competing energies modulate the critical behavior properties of FGT. Additionally, since FGT is a vdW magnet with an easy axis along the crystal's c-axis, we explore how the critical behavior properties modulate with applied $B$ directions. 

\subsection{Order of phase transition}

\begin{figure}[h!]
    \centering
    \includegraphics[width=\linewidth]{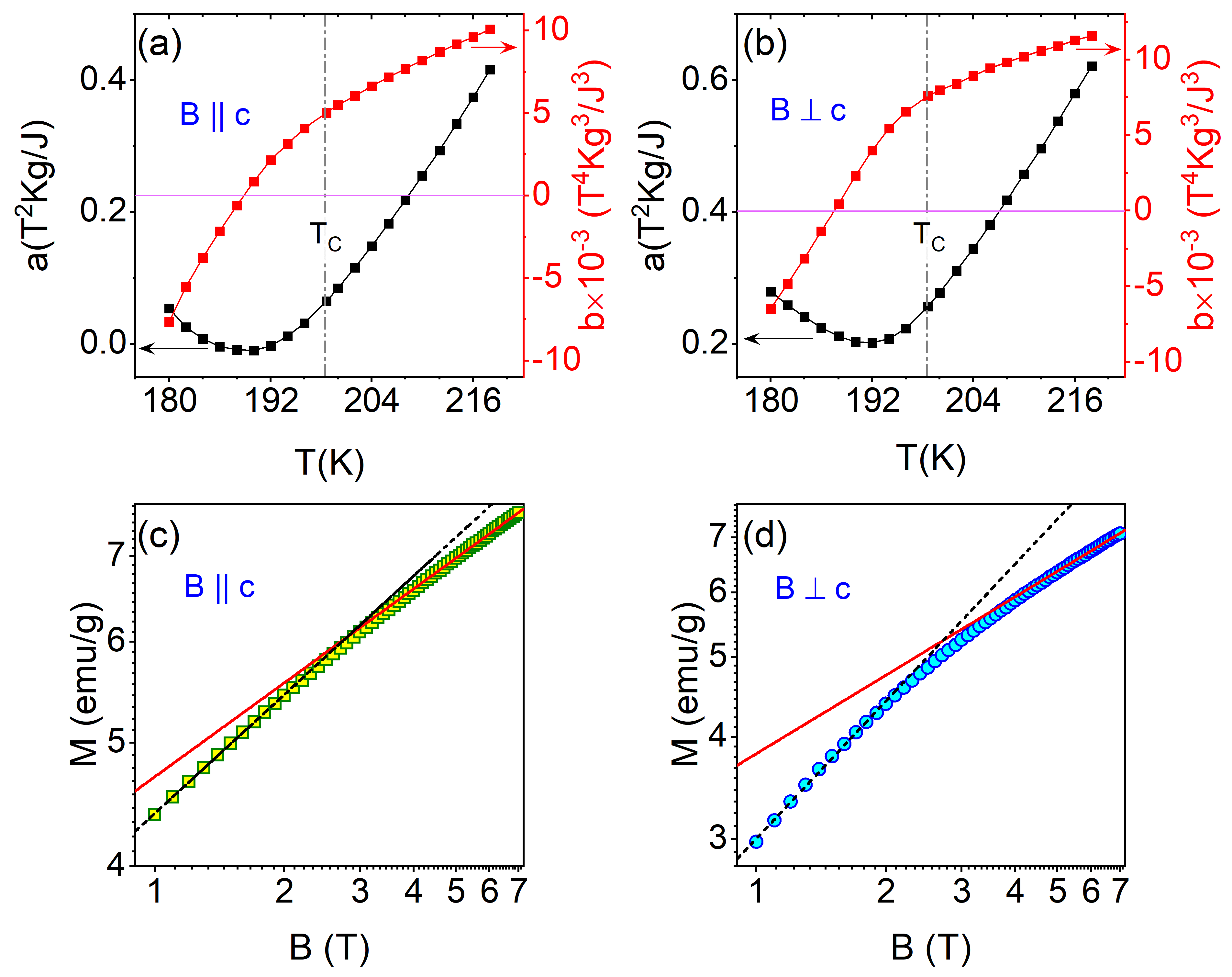}
    \caption{The variation of the Landau coefficients $\mathrm{a(T)}$ and $\mathrm{b(T)}$, with temperature, estimated from the fit to the $M$-$H$ isotherms using Eq. \ref{LandauCoeff} for (a) $B \parallel \mathrm{c}$ and (b) $B \perp \mathrm{c}$. $\mathrm{b(T)}>0$ at $T_{\mathrm{C}}$, indicating second-order phase transition. The log-log plot of the M-H isotherms at $T_{\mathrm{C}}$ for (c) $B \parallel \mathrm{c}$ and (d) $B \perp \mathrm{c}$. The dashed black lines, in the respective figures, are the linear fits to the log-log plot of the $M$-$H$ isotherms for $1\,\mathrm{T}\leq B \leq 2\,\mathrm{T}$. The solid red lines, in the respective figures, are the linear fits to the log-log plot of the $M$-$H$ isotherms for $6\,\mathrm{T}\leq B \leq 7\,\mathrm{T}$.}
    \label{SOPT_MH_Tc}
\end{figure}

The primary test for the study of critical behavior is to check whether the system shows a second-order phase transition (SOPT) at $T_{\mathrm{C}}$. There are several methods to confirm SOPT at $T_{\mathrm{C}}$ in a magnetic system~\cite{Law2018, Franco2006, toledano1987landau}. To determine the order of the phase transition, we apply Landau phenomenological theory~\cite{toledano1987landau}, which states the following. The free energy of a magnetic system, under $B$, can be expressed as the power-law variation of the order parameter $M$ as
\begin{equation}
\begin{split}
 F(M,T)= & F(0) + \frac{a(T)}{2}M^{2} + \frac{b(T)}{4}M^{4} + \frac{c(T)}{6}M^{6} +\\
 & \dots -BM,
 \label{Landau}
 \end{split}
\end{equation}
where $a(T)$, $b(T)$, and $c(T)$ are the Landau coefficients. The variation of $a(T)$, $b(T)$, and $c(T)$ with $T$ depicts the physical characteristics of the magnetic system under investigation. At equilibrium, $\mathrm{dF/dM=0}$, resulting
\begin{equation}
 \mu_{0}H = a(T)M + b(T)M^{3} + c(T)M^{5} + \dots\,.
 \label{LandauCoeff}
\end{equation}
Here, $a\propto (T-T_{\mathrm{C}})$ and $b$ changes its sign at $T=T_{\mathrm{0}}$. A system will show SOPT if $b\geq 0$ at $T=T_{\mathrm{C}}$ and a first-order phase transition if $b<0$ at $T=T_{\mathrm{C}}$. We fit the $M$-$H$ isotherms [Figs.\,\ref{MT_MH_FGT}(c) and \ref{MT_MH_FGT}(d)] using Eq. \ref{LandauCoeff} by restricting it up to $\mathrm{M^5}$~\cite{SM}. We observe $b>0$ [Figs. \ref{SOPT_MH_Tc}(a) and \ref{SOPT_MH_Tc}(b)] at $T=T_{\mathrm{C}}$ for both $B\parallel c$ and $B\perp c$, thus confirming SOPT in FGT. Below, we explore how the coexisting itinerant and localized electrons, FM and AFM interactions, anisotropy, and other competing energies dictate the critical behavior properties of FGT with $\mathrm{B}$. 

\subsection{Field-dependent critical behavior of FGT}
\subsubsection{The Necessity of Field-Dependent Critical Analysis}

\begin{figure}[h!]
    \centering
    \includegraphics[width=0.75\linewidth]{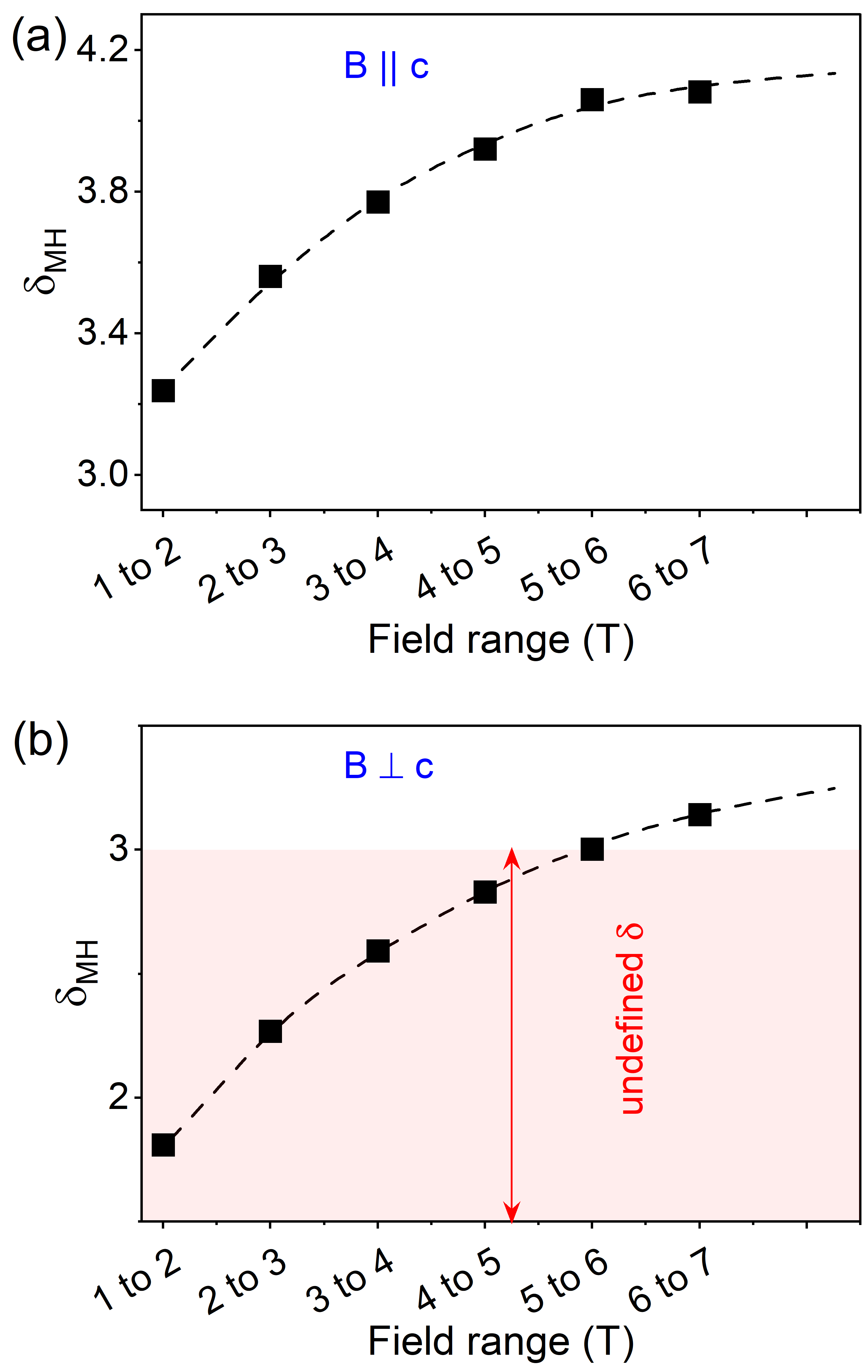}
    \caption{The $\delta_{MH}$ values estimated for different magnetic field ($B$) ranges as mentioned when the magnetic $B$ is applied (a) parallel ($B\parallel c$) and (b) perpendicular ($B\perp c$) to the crystal's c-axis.}
    \label{delta_ANS}
\end{figure}

Recently, Chauhan \textit{et al.}~\cite{chauhan2026magnetic} proposed two methodologies to investigate the $B$-dependent critical behavior of a magnetic system. The log-log plots of the $M$-$H$ isotherms at $T=T_{\mathrm{C}}$ [Figs. \ref{SOPT_MH_Tc}(c) and \ref{SOPT_MH_Tc}(d)] do not show linear variation, confirming the need for a $B$-dependent critical behavior study of the FGT. Following the reported methodologies, we take a fixed temperature range of $\sim 9.4\,\%$ of $T_{\mathrm{C}}$ for both the OOP and IP $M$-$H$ isotherms. We estimate $B$-dependent $\delta_{MH}$ for both OOP [Fig. \ref{delta_ANS}(a)] and IP [Fig. \ref{delta_ANS}(a)] $M$-$H$ isotherms, using the relation
\begin{equation}
    M=A_0 B^{\frac{1}{\delta}},\, \text{at}\, T=T_{\mathrm{C}}, 
    \label{delta_MH}
\end{equation}
where $A_0$ is the proportionality constant. In the measured $B$ range from $\mathrm{1}-\mathrm{7\,T}$, the estimated $\delta_{MH}$ values vary between $\mathrm{3}$ and $\mathrm{5}$ when $B\parallel c$ [Fig. \ref{delta_ANS}(a)], indicating that the FGT spins should reside in three-dimensional ($\mathrm{d=3}$) space. Theoretically, $3\leq\delta < 5$ for $\mathrm{d=3}$ and $\delta = 15$ for $\mathrm{d=2}$~\cite{GarciaGonzalo2003, GarciaMarquesGonzalo2012}. However, when $B\perp c$, the determined $\delta_{MH}$ values at low $B$ [Fig. \ref{delta_ANS}(b)] violate the theoretical developments. This is evidence that the magnetocrystalline anisotropy plays a key role in defining the critical behavior of a magnetic system.  When $B\parallel c$ [Fig. \ref{delta_ANS}(a)], the $\delta_{MH}$ seems to attain saturation rapidly, though the exact saturation point lies beyond the applied $B$ range. Conversely, when $B\perp c$ [Fig. \ref{delta_ANS}(b)], a much larger $B$ is required to reach the saturation of $\delta_{MH}$. 
\begin{figure}[htb!]
    \centering
    \includegraphics[width=0.8\linewidth]{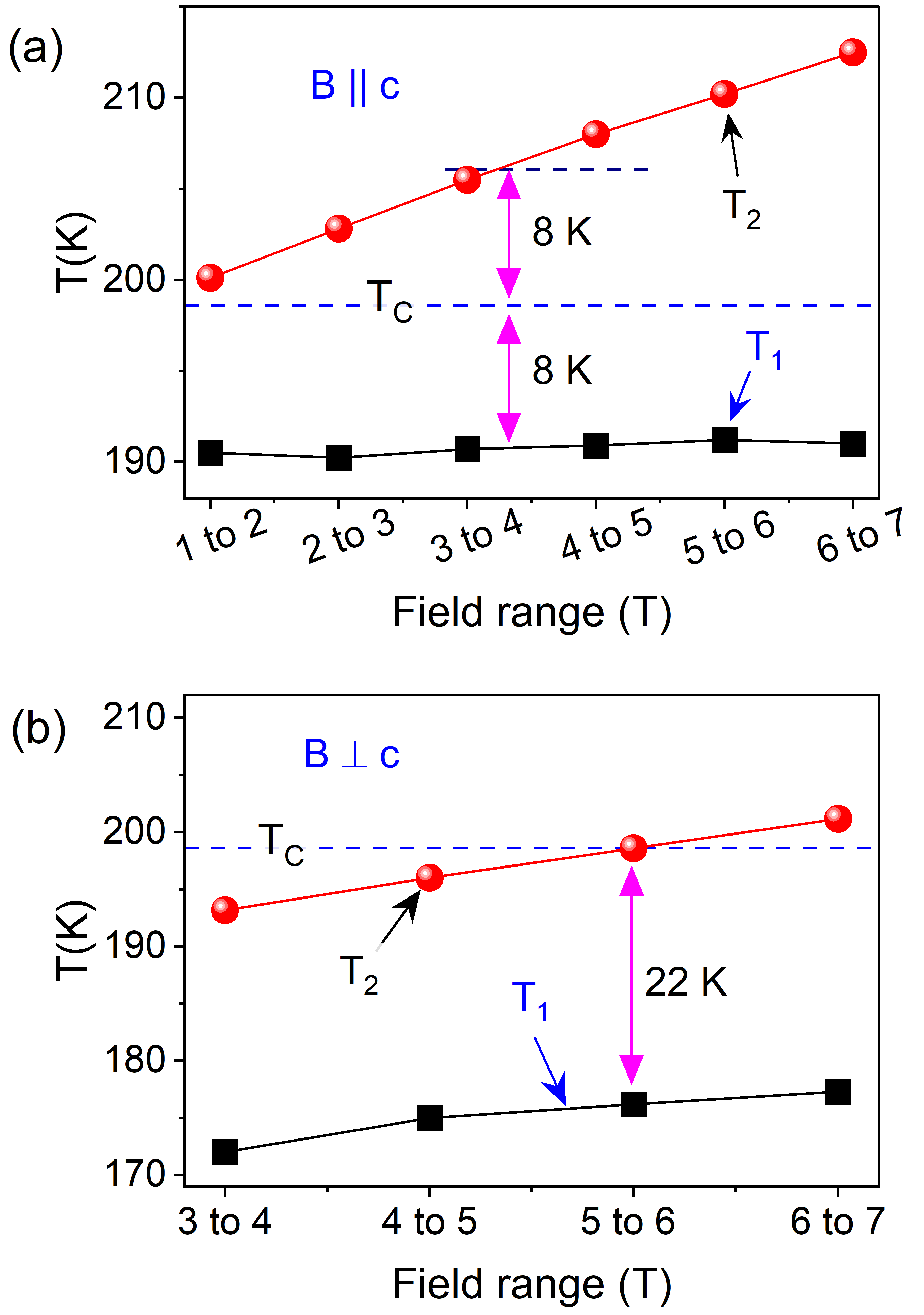}
    \caption{The variation of crossover temperatures $\mathrm{T_1}$ and $\mathrm{T_2}$ with field when (a) $B\parallel c$ and (b) $B\perp c$. The $\mathrm{T_1}$ and $\mathrm{T_2}$ values are estimated from the crossover of the intercepts, which were estimated from the linear fit to the modified Arrott plots constructed using the critical exponent values associated with SR standard universality classes and the LR mean-field universality class, respectively.}
    \label{ROI_both}
\end{figure}

\subsubsection{Estimating the range of exchange interaction: the intercept analysis}

In 1972, Fisher \textit{et al.} \cite{PhysRevLett.29.917} first proposed the relation of $\gamma$ with $\sigma$, $d$ and $n$ as
\begin{equation}
\begin{split}
\gamma = 1 + \displaystyle\frac{4}{d}\Bigg(\displaystyle\frac{n+2}{n+8}\Bigg)\Delta{\sigma}+\Bigg(\displaystyle\frac{8(n+2)(n-4)}{d^2(n+8)^2}\Bigg)
\\\times\left[{1+\displaystyle\frac{2G\left(\displaystyle\frac{d}{2}\right)(7n+20)}{(n-4)(n+8)}}\right]\Delta{\sigma^2}+...,
\end{split}
\label{WilsonFisher}
\end{equation}
where $\Delta{\sigma}$ = $\left(\sigma - {\displaystyle\frac{d}{2}}\right)$ and $G\left(\displaystyle\frac{d}{2}\right) = 3 - \displaystyle\frac{1}{4}\left(\displaystyle\frac{d}{2}\right)^2$. Equation \ref{WilsonFisher} is used to estimate $\sigma$, which eventually evaluates $J(r)\sim r^{-(d+\sigma)}$ emerging in a magnetic system. The proposed relation is exactly valid for a trivial magnetic system with LR interaction; however, it has been used for systems with SR interaction as well~\cite{liu2017critical, Liu2017,AnkitaPhysRevB.109.L020407, TIWARI2021168020, tiwari2020ising}. It is stated that a system will show LR interaction if $\sigma \leq 1.5$ and SR (SR) interaction if $\sigma \geq 2.0$. However, there is a speculative explanation regarding the range of interaction if $\sigma$ lies between $1.5$ and $2.0$. Several groups treat the magnetic system as weakly LR for $\mathrm{1.5}<\sigma< \mathrm{2.0}$~\cite{liu2017critical, Liu2017, BHATTACHARYA2023170274, TIWARI2021168020}. The recent investigation claims that the Fisher \textit{et al.} relation (Eq. \ref{WilsonFisher}) may not yield the exact value of $\sigma$ for a nontrivial magnetic system~\cite{chauhan2022different, chauhan2026magnetic}. In this case, no method exists other than the recently proposed qualitative approach to identify the range of the exchange interaction by estimating the intercepts of the linear fits to modified Arrott plots (MAPs) constructed using the CE values of the standard UCs~\cite{chauhan2026magnetic}. The intercept versus temperature plots (see SM~\cite{SM}) show crossover at temperature $T$ for a set of CE values belonging to a standard UC. We define $\mathrm{T_1}$ as the crossover temperature belonging to SR standard UCs, viz., 3D Heisenberg, 3D XY, and 3D Ising, and $\mathrm{T_2}$ as the crossover temperature associated with the LR mean-field UC. 

\begin{figure*}
    \centering
    \includegraphics[width=0.8\linewidth]{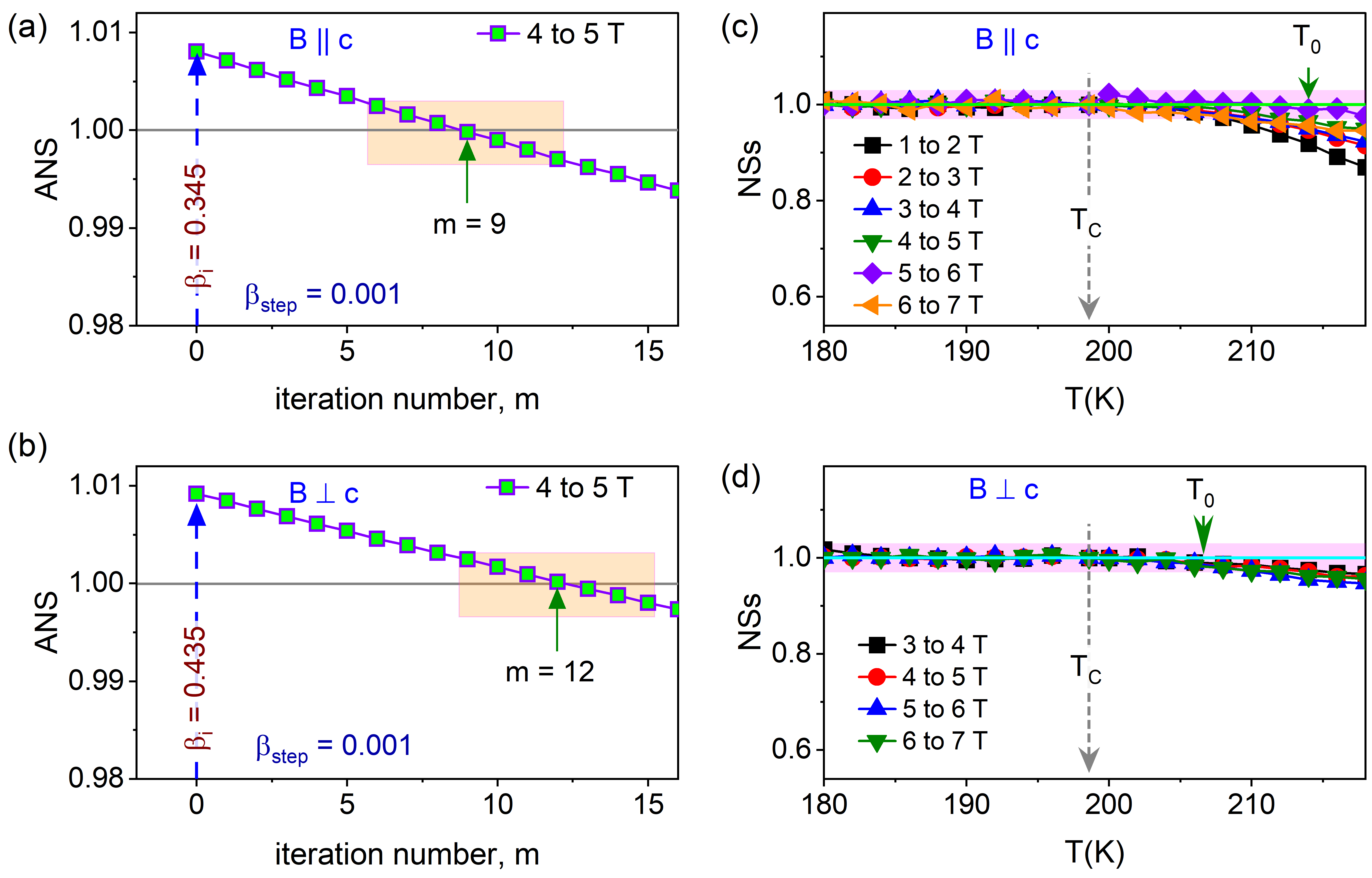}
    \caption{The average of normalized slope (ANS) plots with temperature for the FGT when (a) $B \parallel c$ and (b) $B \perp c$. The ANS is estimated using the relation \ref{ANS} as discussed in the main text. (c) and (d) the estimated normalized slopes (NSs) from the linear fit to modified Arrott plots constructed using the final critical exponent values obtained using the ANS method for $B \parallel c$ and $B \perp c$, respectively. The $\mathrm{T_0}$ in the respective figure is the temperature value up to which the NSs remain close to 1 above $T_{\mathrm{C}}$ for the $B$ range of $\mathrm{6}$ to $\mathrm{7\,T}$.}
    \label{ANS_NSs}
\end{figure*}

\begin{table*}
	\centering
	\caption{The list of estimated field-dependent CE values for $\mathrm{Fe_3GeTe_2}$ in this work, along with reported literature on the CE values for $\mathrm{Fe_3GeTe_2}$ and the defined UCs. The symbols have their usual meaning as discussed in the main text. MAP: modified Arrott plot, ANS: average of normalized slope.}
\begingroup
\renewcommand{\arraystretch}{1.5} 
	\begin{tabular}{lllcllllll}
		\hline
        \hline
		\multirow{2}{*}{Material/Models} & \multirow[c]{2}{*}{Reference}& \multirow[c]{2}{*}{Method} & {Field range} & {$T_{\rm{C}}$} & {} & \multicolumn{4}{c}{Critical Exponents} \\ 
		\cline{6-10}
		{} & {} & {} & {($\rm{T}$)} & {($\rm{K}$)} & {$\delta$ ($\delta_{MH}$)} &{$\beta$} & {$\gamma$} & {$\sigma_{eff}$} & {$\sigma$} ($d:n$) \\
		\hline
		
		{Mean-field (MF)} & \cite{PhysRevB.65.214424, KAUL19855, PhysRevB.21.3976} & Theory & \multicolumn{2}{c}{} & {3.000} & {0.500} & {1.000} & {1.500} & {1.500} \\
        
		{Tricritical MF} & \cite{PhysRevB.65.214424, PhysRevB.21.3976} & Theory & \multicolumn{2}{c}{} & {5.000} & {0.250} & {1.000} & {} & {1.000} \\
		
		{3D Ising} & \cite{PhysRevB.65.214424, KAUL19855, PhysRevB.21.3976} & Theory & \multicolumn{2}{c}{} & {4.800} & {0.325} & {1.237} & {1.967} &{1.967} \\
		
		{3D XY} & \cite{PhysRevB.65.214424, KAUL19855, PhysRevB.21.3976} & Theory & \multicolumn{2}{c}{} & {4.800} & {0.345} & {1.316} & {1.968} & {1.963} \\
		
		{3D Heisenberg} & \cite{PhysRevB.65.214424, KAUL19855, PhysRevB.21.3976} & Theory & \multicolumn{2}{c}{} & {4.800} & {0.365} & {1.386} & {1.965} & {1.965} \\

        \multirow[t]{4}{*}{$\mathrm{Fe_3GeTe_2}$} & \multirow[t]{4}{*}{\cite{AnkitaPhysRevB.109.L020407}} & {MAP} & {0.25$-$0.75} & {{223}$^{MT}$} & {3.000} & {0.500} & {1.00} & {} & {1.000} ($2:3$) \\

        \multirow[t]{4}{*}{$\mathrm{Fe_3GeTe_2}$} & \multirow[t]{4}{*}{\cite{AnkitaPhysRevB.109.L020407}} & {MAP} & {0.75$-$1.5} & {} & {3.292} & {0.439} & {1.01} & {} & {1.011} ($2:3$) \\

        \multirow[t]{4}{*}{$\mathrm{Fe_3GeTe_2}$} & \multirow[t]{4}{*}{\cite{AnkitaPhysRevB.109.L020407}} & {MAP} & {1.5$-$3.0} & {} & {3.684} & {0.388} & {1.05} & {} & {1.050} ($2:3$) \\

        \multirow[t]{4}{*}{$\mathrm{Fe_3GeTe_2}$} & \multirow[t]{4}{*}{\cite{AnkitaPhysRevB.109.L020407}} & {MAP} & {3.0$-$5.0} & {} & {3.863} & {0.384} & {1.09} & {} & {1.089} ($2:3$) \\

        \multirow[t]{4}{*}{$\mathrm{Fe_3GeTe_2}$} & \multirow[t]{4}{*}{\cite{liu2017critical}} & {MAP} & {0$-$8} & {{215}$^{MT}$} & {4.261(9)} & {0.327(3)} & {1.079(5)} & {} & {} \\

        \multirow[t]{4}{*}{$\mathrm{Fe_3GeTe_2}$} & \multirow[t]{4}{*}{\cite{Liu2017}} & {MAP} & {0.3$-$5} & {{152}$^{MT}$} & {4.50(1)} & {0.372(4)} & {1.265(15)} & {} & {} \\
		
		\multirow[t]{6}{*}{$\mathrm{Fe_3GeTe_2}$ ($B\parallel c$)} & \multirow[t]{6}{*}{This work} & \multirow[t]{6}{*}{ANS} & 1$-$2  & \multirow[t]{6}{*}{{198.6}$^{MT}$} & {$3.237(20)$} & {0.432(3)} & {0.967(6)} & {1.584(61)} & {1.420(16)} ($3:1$) \\
		
		{} & {} & {} & 2$-$3 & {} & {$3.560(13)$} & {0.394(3)} & {1.009(8)} & {1.684(79)} & {$1.520(17)$} ($3:1$) \\
		
		{} & {} & {} & 3$-$4 & {} & {$3.771(16)$} & {0.365(3)} & {1.011(8)} & {1.742(84)} & {$1.524(17)$} ($3:1$) \\
		
		{} & {} & {} & 4$-$5 & {} & {$3.921(14)$} & {0.354(3)} & {1.034(9)} & {1.781(92)} & {$1.572(18)$} ($3:1$) \\
		
		{} & {} & {} & 5$-$6 & {} & {$4.060(15)$} & {0.350(3)} & {1.071(11)} & {1.814(106)} & {$1.642(20)$} ($3:1$) \\
		
		{} & {} & {} & 6$-$7 & {} & {$4.082(16)$} & {0.330(3)} & {1.017(10)} & {1.819(105)} & {$1.537(21)$} ($3:1$) \\
		
		\multirow[t]{6}{*}{$\mathrm{Fe_3GeTe_2}$ ($B\perp c$)} & \multirow[t]{6}{*}{This work} & \multirow[t]{6}{*}{ANS} & 3$-$4  & \multirow[t]{6}{*}{{198.6}$^{MT}$} & {$2.592(14)$} & {0.500(3)} & {0.796(5)} & {1.330(49)} & {NA} \\
		
		{} & {} & {} & 4$-$5 & {} & {$2.831(11)$} & {0.447(3)} & {0.819(5)} & {1.434(54)} & {NA} \\
		
		{} & {} & {} & 5$-$6 & {} & {$3.003(10)$} & {0.411(3)} & {0.823(6)} & {1.501(66)} & {NA} \\
		
		{} & {} & {} & 6$-$7 & {} & {$3.141(9)$} & {0.392(3)} & {0.839(6)} & {1.551(68)} & {NA} \\
		\hline
        \hline
	\end{tabular}
    \endgroup
	\label{Table_Critical exponents}
\end{table*}

Figure \ref{ROI_both} shows the variation of $\mathrm{T_1}$ and $\mathrm{T_2}$ with the $B$ range that was used to estimate the values of $\mathrm{T_1}$ and $\mathrm{T_2}$~\cite{SM}. We observe an almost constant shift of $\mathrm{T_1}$ from $T_\mathrm{C}$ irrespective of the used $B$ range for both $B \parallel c$ [Fig. \ref{ROI_both}(a)] and $B \perp c$ [Fig. \ref{ROI_both}(b)], however the distance of $\mathrm{T_2}$ from $T_\mathrm{C}$ varies with increasing $B$. This is an indication that the range of exchange also depends on $B$. At low $B$, the FGT exhibits dominant LR exchange interaction, which is suppressed by increasing $B$. The variation of $\mathrm{T_1}$ and $\mathrm{T_2}$, thus, indicates that the LR interaction becomes weaker with $B$ when $B \parallel c$ and remains dominant when $B \perp c$. These observations confirm the role of competing energies in determining the critical behavior of FGT. The following two points justify the observed variation of LR behavior. (i) The itinerant electrons cause LR Ruderman-Kittel-Kasuya-Yosida (RKKY) interaction which suppresses with $B$. (ii) The existing double exchange mechanism between $\mathrm{Fe^{3+}}$ and $\mathrm{Fe^{2+}}$ ions also leads to LR interaction due to an emerging percolation mechanism, which also suppresses with $B$. The $B$-direction-dependent qualitative behavior of the existing range of exchange interaction indicates the strong effect of uniaxial magnetocrystalline anisotropy (easy axis) and thus the role of dipolar energy in modulating the critical behavior of a magnetic system. We thus argue that the $B$-direction-dependent critical behavior will be strong in vdW magnets. Below, we estimate $\beta$, $\gamma$, and subsequently $\sigma_{eff}$ to quantitatively evaluate the range and type of exchange interaction exhibited in FGT as a function of the applied $B$ direction using the proposed ANS analysis method~\cite{chauhan2026magnetic}. 

\subsubsection{Estimating the field-dependent critical exponents: the ANS method}

The proposed ANS method~\cite{chauhan2026magnetic} states that the estimation of appropriate $\beta$ and $\gamma$ values for $T\leq T_{\mathrm{C}}$ is sufficient to explore the critical behavior of a magnetic system. Therefore, hereafter we only discuss the critical behavior of FGT using the estimated $\beta$ and $\gamma$ values for $T\leq T_{\mathrm{C}}$. The proposed ANS method is as follows. First, estimate $\delta_{MH}$ for a fixed $B$-range using Eq. \ref{delta_MH}. Then, correlate $\beta$ and $\gamma$ with $\delta_{MH}$ using the Widom relation as $\gamma=\beta\times(\delta_{MH}-1)$. Now vary $\beta$ as
\begin{equation}
	\beta_m=\beta_i\pm m \times\beta_{step},
\label{beta_m}
\end{equation}
where $\beta_i$ is the initially selected value, $\beta_{step}$ is the step increment in $\beta$ for each iteration and $m$ is the positive integer representing the iteration number. To start with, $\beta_i$ can be chosen either as a standard theoretical value corresponding to any standard UC or as a random value, with the condition that the chosen set of $\beta$ and corresponding $\gamma$ leads to linear behavior of the constructed MAPs, see \cite{SM}. Thus, for a selected $\beta_m$, determine the corresponding $\gamma_m$ using Eq. \ref{beta_m} as 
\begin{equation}
	\gamma_m=\beta_m\times(\delta_{MH}-1).
	\label{gamma_m}
\end{equation}
Now, construct the MAPs using the above correlated $\beta_m$ (Eq. \ref{beta_m}) and $\gamma_m$ (Eq. \ref{gamma_m}) values. Next, estimate normalized slopes (NSs) using the slopes of the linear fit to the MAPs. Determine the ANS using the estimated NS values as
\begin{equation}
	ANS=\frac{\sum_{i=1}^{N} (NS)_i}{N},
	\label{ANS}
\end{equation}
where $i$ denotes the $M-H$ isotherm at a given $T$, $\sum_{i=1}^{N} (NS)_i$ is the sum of the NSs, and $N$ is the total number of isotherms used for the estimation. Thus, for a selected $\beta_m$ and corresponding $\gamma_m$, we get an ANS value. Figures \ref{ANS_NSs} (a) and \ref{ANS_NSs}(b) show the representative ANS plots with $n$ for the $B$ range of $\mathrm{4}-\mathrm{5\,T}$ when $B \parallel c$ and $B \perp c$, respectively. We vary $\beta_m$ and the corresponding $\gamma_m$ in such a way that the ANS $\rightarrow1$. We stop the increment or decrement of $\beta_m$ and the corresponding $\gamma_m$ when ANS$=1$ within the error limit. The error estimation varies depending on the selected $\beta_{step}$. For $\beta_{step}=0.001$, ANS$=1.00\pm0.005$ [see shaded regions in Figs. \ref{ANS_NSs} (a) and \ref{ANS_NSs}(b)] is best to get appropriate values of $\beta$ and $\gamma$. Following the above-discussed processes, we estimate the $B$-dependent $\beta$ and $\gamma$ values (see Table \ref{Table_Critical exponents}) for both $B \parallel c$ and $B \perp c$. 

The simplest and effective way to test the robustness of the estimated $\beta$ and $\gamma$ values is the estimation of NSs for the $M$-$H$ isotherms recorded in the vicinity of $T_{\mathrm{C}}$. The $\beta$ and $\gamma$ values are considered appropriate if the NS values remain close to 1 irrespective of the temperature. As shown in Figs. \ref{ANS_NSs} (c) and \ref{ANS_NSs}(d), the NS values remain close to $\mathrm{1}$ irrespective of the temperature for $T\leq T_{\mathrm{C}}$. The NS $\sim \mathrm{1}$ in the vicinity of $T_{\mathrm{C}}$ confirms that the critical behavior remains the same on either side of $T_{\mathrm{C}}$, however, one needs to vary the temperature range for $T>T_{\mathrm{C}}$ depending on the selected $B$ range. Below, we discuss the role of the competing energies and the effect of $B$-direction on the critical behavior of FGT.

\subsubsection{The critical behavior of FGT: the role of competing energies}

\begin{figure*}
    \centering
    \includegraphics[width=0.85\linewidth]{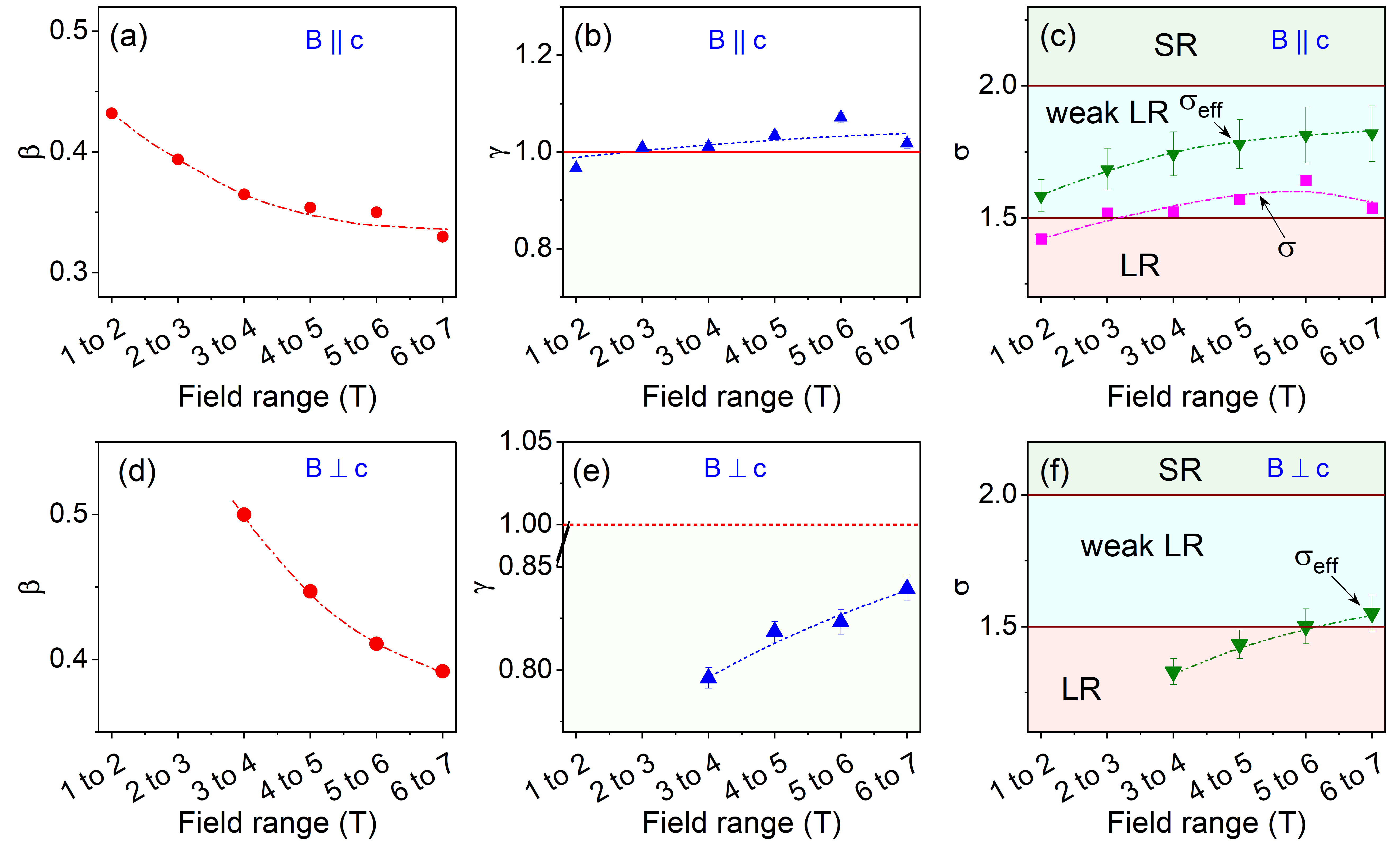}
    \caption{The variation of (a) $\beta$, (b) $\gamma$ and (c) $\sigma_{eff}$ with the applied $B$ range when $B \parallel c$. The $\beta$ converges 3D Ising universality class at higher $B$, while $\gamma$ remains close to $\mathrm{1}$. The variation of $\sigma_{eff}$ indicates the modulation of LR interaction from strong to weak with increasing $B$. The variation of (d) $\beta$, (e) $\gamma$ and (f) $\sigma_{eff}$ with the applied $B$ range when $B \perp c$. The $\beta$ gradually decreases below $\mathrm{0.400}$ and does not converge in the measured $B$ range. The $\gamma$ slowly increases, but does not reach $\mathrm{1}$ and beyond in the measured $B$ range, indicating a strong effect of anisotropy energy. $\sigma_{eff}$ gradually increases but remains close to $\mathrm{1.5}$ indicating dominant LR interaction.}
    \label{CA_FGT}
\end{figure*}

When $B \parallel c$, the $\beta$ values [Fig. \ref{CA_FGT}(a) and Table \ref{Table_Critical exponents}] vary from $\sim\mathrm{0.43}$ to $\sim\mathrm{0.33}$ with increasing $B$ while the $\gamma$ values remains close to $1$ within the measured $B$ range. To explain this variation, we delve into the properties of FGT. In FGT, at least five energy parameters, viz., double exchange, direct exchange, RKKY, dipolar, and anisotropy energies, modulate the phase transition and thus the critical behavior~\cite{yi2016competing, Deng2018, Tan2018,  Fei2018, liu2017critical, Liu2017,10.1063/1.4961592, AnkitaPhysRevB.109.L020407, Kim2026, yadav2026intrinsic, zhang2018emergence, sharma2023non}. Since FGT is a nontrivial magnetic system, we get the effective CE values using the ANS method. We thus argue that the effective $\beta \sim 0.43$ at low $B$ belongs to the mean-field UC. With increasing $B$, this $\beta$ value is suppressed due to the modulation of the competing energies and ends up belonging to the 3D Ising UC. Since the estimated $\beta$ value is effective, it will be hard to observe its convergence to a standard UC. Following the recent investigation~\cite{chauhan2026magnetic}, we restate the above explanations as follows.

When $B \parallel c$, the double exchange and RKKY interactions cause LR ordering, resulting in $\gamma\sim1$ with no sign of a monotonic increase within the measured $B$ range. The anisotropy and dipolar energies restrict $n$, which is clearly visible in the high-$B$ critical behavior (Table \ref{Table_Critical exponents}). Taking into consideration only the $B$-dependent $\beta$, we find that $n=3$ ($\beta=0.365$) in the range of $3-4$ T and $\mathrm{n} \rightarrow1$ ($\beta\rightarrow 0.325$) in the high-$B$ range. For $\beta>>0.365$, the $n$ values have no meaning as the system exhibits LR interaction. The values of $n$ change due to the suppression of LR interactions with $B$. These findings directly support the evaluated $B$-dependent variation of the range of exchange interactions obtained using intercept analysis; see Fig. \ref{ROI_both}.

We now attempt to quantitatively determine the range of exchange interactions by estimating $\sigma_{eff}$ and $\sigma$ (Eq. \ref{WilsonFisher}). One can estimate $\sigma_{eff}$ by treating the Rushbrooke inequality as an equality, $\alpha+2\beta+\gamma=2$. Furthermore, $\gamma$ is related to $\sigma$ and $\nu$ by $\sigma=\frac{\gamma}{\nu}$, while $\alpha=2-\nu d$. Combining these relations, we estimate
\begin{equation}
    \sigma_{eff}=\frac{\gamma d}{2\beta+\gamma}.
\label{Sigma_eff}
\end{equation}
Equation \ref{Sigma_eff} does not involve $n$, and the value of $d$ is also unknown. However, based on the estimated $\delta_{MH}<5.0$ [Fig. \ref{delta_MH}(a)], we assume $d=3$~\cite{PhysRevB.65.214424, KAUL19855}. Thus, by substituting $d=3$ along with the $\beta$ and $\gamma$ values, we estimate the $B$-dependent $\sigma_{eff}$, see Table \ref{Table_Critical exponents}. As shown in Fig. \ref{CA_FGT}(c), at low $B$, $\sigma_{eff}$ indicates the stabilization of the dominant LR interaction, which weakens with increasing $B$. This behavior is fully justified by the intercept analysis [Fig. \ref{ROI_both}(a)].

Furthermore, one can estimate $\sigma$ by substituting the $d$, $n$ and $\gamma$, associated with localized spins, into Eq. \ref{WilsonFisher}. Usually, this relation is used to determine $d$, $n$ and $\sigma$. Here, the estimated $n$ for a particular $d$ will represent the UC of the magnetic system under investigation~\cite{Kouvel1964, Zhu2023, Gupta2018, Tateiwa2014, Bhattacharyya2011, chauhan2022different}. For FGT, the determined $\sigma$ (Table \ref{Table_Critical exponents}) is closer to $\sigma_{eff}$ for $\mathrm{d=3}$ and $\mathrm{n=1}$, indicating that the critical behavior of the localized spins belongs to 3D Ising UC. This is further supported by the intrinsic anisotropy and dipolar energies, which restrict $n$ of the local electronic moments.

When the $B$ is applied perpendicular to the crystal's c-axis (i.e., in the IP direction), the obtained CE values for FGT [Figs. \ref{CA_FGT}(d) to \ref{CA_FGT}(f) and Table \ref{Table_Critical exponents}] exhibit unusual characteristics that are not supported by existing theories~\cite{stanley1999scaling, fisher1974renormalization, chauhan2026magnetic, Liu2017, liu2017critical}. As shown in Fig. \ref{CA_FGT}(d), the $\beta$ value does not converge within the measured $B$ range. However, the variation of $\beta$ from $\sim0.5$ at low $B$ to $\sim0.4$ at high $B$ indicates the domination of the LR interactions within the measured $B$ range. The estimated $\gamma$ value (for $T<T_{\rm{C}}$) [Fig. \ref{CA_FGT}(e)] monotonically increases with $B$ but remains strictly below the theoretical threshold of $\geq1$ across the measured window up to $\rm{7\,T}$. While standard scaling laws mandate $\gamma \geq1$ for isotropic phase transitions~\cite{PhysRevB.65.214424, KAUL19855, PhysRevB.21.3976}, such sub-unity values [Fig. \ref{CA_FGT}(e)] can be understood as follows. We estimate $\gamma$ using the $M$-$H$ isotherms for $T\leq T_{\rm{C}}$, whereas typically reported $\gamma$ values are extracted from the initial inverse susceptibility $\chi_0^{-1}(T)$ for $T\geq T_{\rm{C}}$ using conventional analysis methods~\cite{ANPhysRevLett.19.786, Kouvel1964, Zhu2023, Gupta2018, Tateiwa2014, Bhattacharyya2011}. The magnetization process is dominated by coherent spin rotation against the anisotropy barrier rather than purely thermodynamic spin fluctuations. This introduces an effective $B$-dependent constraint (on $n$), suppressing the divergence of $\chi$ near $T_{\rm{C}}$ and artificially truncating the scaling of the MAPs. Consequently, $\gamma<1.0$ highlights the limitations of equilibrium scaling models when treating strongly anisotropic systems in their hard-axis (IP) orientation, demonstrating that the apparent critical parameters (Table \ref{Table_Critical exponents}) are effective rather than intrinsic thermodynamic exponents. Moreover, in an anisotropy-driven magnetic system, $\gamma$ can differ on either side of $T_{\rm{C}}$, typically taking a lower value for $T < T_{\rm{C}}$~\cite{PhysRevLett.115.200601}. These discussions indicate that the obtained sub-unity value of $\gamma$ [Fig. \ref{CA_FGT}(e)] may actually reflect the intrinsic physical characteristics of $\chi$ for $T\leq T_{\rm{C}}$ in highly anisotropic magnetic systems. This behavior is well supported in FGT, which exhibits strong uniaxial magnetocrystalline anisotropy~\cite{yi2016competing, Deng2018, Tan2018,  Fei2018, liu2017critical, Liu2017,10.1063/1.4961592, AnkitaPhysRevB.109.L020407,Kim2026, yadav2026intrinsic}. 

Moreover, the $\delta_{MH}$ achieves the theoretically predicted value beyond $B=\rm{5\,T}$~\cite{PhysRevB.65.214424, KAUL19855, PhysRevB.21.3976}. This critical $B$ value signifies the threshold where the Zeeman energy overcomes the magnetocrystalline anisotropy energy for $B \perp c$. Nevertheless, the MAPs~\cite{SM} constructed for the $M$-$H$ isotherms for $T\leq T_{\mathrm{C}}$ exhibit quasi-linear behavior in this $B$ range. We further verify the quasi-linear behavior of the MAPs by demonstrating a horizontal collapse of all isotherms for the first derivative of the MAPs~\cite{AnkitaPhysRevB.109.L020407, SM}. 

These observations—specifically observing $\gamma<1.0$~\cite{chauhan2026magnetic, zhang2013influence} despite the MAPs showing quasi-linear behavior and NSs$\sim1$ pose serious questions about the complete understanding of the critical behavior of nontrivial magnetic systems. The lack of complete convergence in $\beta$ and $\gamma$ values for $B\perp c$ within the $1–7\, \rm{T}$ window suggests that the applied Zeeman energy has not yet fully overcome the intrinsic magnetocrystalline anisotropy energy barrier in FGT. At much higher $B$ ($>7\,\rm{T}$), where the Zeeman energy completely dominates over the anisotropy energy, the local spins are expected to align fully along the hard axis. In this high-$B$ limit, the critical parameters should undergo a crossover, recovering a physically valid regime where $\beta$ and $\gamma$ ultimately converge to standard UC. Future high-$B$ magnetization experiments are therefore necessary to completely map the $B$-dependent phase diagram and confirm the isotropic nature of the intrinsic CEs in this class of vdW ferromagnets.

The estimated $\sigma$ [Fig. \ref{CA_FGT}(f)] indicates dominating LR interactions within the measured $B$ range up to $B=\mathrm{7\,T}$ for the $B\perp c$ configuration. However, for such low values of $\gamma<1.0$, Eq. \ref{WilsonFisher} yields imaginary values for $\sigma$. Specifically, Eq. \ref{WilsonFisher} can only yield real values for $\sigma$ when $\gamma>\mathrm{0.855}$ (for $d=3$, $n=3$), $\gamma>\mathrm{0.863}$ (for $d=3$, $n=2$), and $\gamma>\mathrm{0.871}$ (for $d=3$, $n=1$). This highlights an additional challenge: the lack of an established methodology to reliably determine $n$ for a nontrivial magnetic system undergoing a continuous phase transition under these specific critical conditions.

\section{Conclusions}
In conclusion, we have presented a comprehensive investigation into the critical phenomena of the vdW ferromagnet $\mathrm{Fe_3GeTe_2}$ by employing the recently developed ANS method~\cite{chauhan2026magnetic}. Our experimental and analytical results reveal that the critical scaling parameters in this
system are fundamentally modulated by both the strength and spatial orientation of $B$, challenging conventional critical behavior investigation.

For the easy-axis configuration ($B\parallel c$), the system undergoes a continuous $B$-driven crossover. At low $B$, the critical behavior is governed by LR magnetic interactions mediated by a competitive interplay of double-exchange mechanisms and itinerant electron (RKKY) magnetism, yielding a mean-field-like exponent ($\beta\sim0.43$, $\gamma\sim1.0$, $\delta_{MH}\sim 3.24$). With increasing Zeeman energy, these LR interactions are systematically suppressed, driving the system to a clean convergence within LR 3D Ising UC ($\beta\rightarrow0.330$, $\gamma\sim1.0$) at high $B$.

Conversely, the hard-axis configuration ($B\perp c$) uncovers an anomalous critical regime characterized by sub-unity susceptibility exponents ($\gamma<1.0$) across the entire measured $B$ window. Although this behavior is theoretically undefined, we propose that $\gamma$ can acquire a sub-unity value in magnetic systems with strong uniaxial magnetocrystalline anisotropy, as observed in FGT.
We further show that the obtained CE values mark a clear crossover where the strong OOP magnetocrystalline anisotropy energy barrier strongly constrains conventional thermodynamic spin fluctuations. In this regime, the magnetization profile is heavily driven by a coherent spin rotation toward the hard plane rather than pure thermodynamic phase transition dynamics, rendering the extracted CE values effective rather than intrinsic exponents. The lack of convergence in $\beta$ and $\gamma$ values in the measured $B$-range suggests that the intrinsic critical behavior of FGT should remain isotropic irrespective of the applied $B$ direction.

Ultimately, this study underscores that evaluating CEs across fixed, broad $B$ windows in strongly anisotropic or low-dimensional magnets can obscure or distort the underlying physical scaling. The success of the $B$-dependent ANS approach in mapping the complex magnetic landscape of $\mathrm{Fe_3GeTe_2}$ establishes it as a vital diagnostic tool for resolving competing interactions in emerging quantum and topological van der Waals materials.

\section{Acknowledgments}
NY is thankful to the Council of Scientific and Industrial Research (CSIR), Government of India, for the research fellowship. NY, HCC and PD acknowledge the Department of Physics, Central Research Facility (CRF) and Nano Research Facility (NRF) of the Indian Institute of Technology (IIT) Delhi, for characterization facilities. PD gratefully acknowledges the partial funding support from the Science and Engineering Research Board (SERB) of the Government of India (grant no. SPR/2021/000762), Indo-French Centre for the Promotion of Advanced Research (IFCPAR/CEFIPRA) through Grant No. 6508-2/2021 and the National Quantum Mission, an initiative of the Department of Science and Technology, Government of India. HCC gratefully acknowledges the Anusandhan National Research Foundation (ANRF), Government of India, for the partial financial support through the National Post-Doctoral Fellowship (N-PDF) (Grant/Sanction No. PDF/2026/005456).

\section*{Data Availability Statement}
The data supporting this study's findings are available from the corresponding author upon reasonable request.


\begin{thebibliography}{99}

\bibitem{stanley1999scaling}
H. E. Stanley,
Scaling, Universality, and renormalization: Three pillars of modern critical phenomena,
\href{https://doi.org/10.1103/RevModPhys.71.S358} {Rev. Mod. Phys. \textbf{71}, S358--S366 (1999)}.


\bibitem{fisher1974renormalization}
M. E. Fisher,
The renormalization group in the theory of critical behavior,
\href{https://doi.org/10.1103/RevModPhys.46.597} {Rev. Mod. Phys. \textbf{46}, 597--616 (1974)}.


\bibitem{phan2010tricritical}
M.-H. Phan, V. Franco, N. S. Bingham, H. Srikanth, N. Hur, and S.-C. Yu,
Tricritical point and critical exponents of La$_{0.7}$Ca$_{0.3-x}$Sr$_x$MnO$_3$ ($x=0$, 0.05, 0.1, 0.2, 0.25) single crystals,
\href{https://doi.org/10.1016/j.jallcom.2010.07.223}{J. Alloys Compd. \textbf{508}, 238--244 (2010)}.


\bibitem{tiwari2020ising}
J. K. Tiwari, H. C. Chauhan, B. Kumar, and S. Ghosh,
3D-Ising like ferromagnetism in skyrmionic-bubbles host infinite-layer La$_{0.825}$Sr$_{0.175}$MnO$_3$ manganite perovskite,
\href{https://doi.org/10.1088/1361-648X/ab6d14}{J. Phys.: Condens. Matter \textbf{32}, 195803 (2020)}.


\bibitem{scheie2022spin}
A. Scheie, P. Laurell, P. A. McClarty, G. E. Granroth, M. B. Stone, R. Moessner, and S. E. Nagler,
Spin-exchange Hamiltonian and topological degeneracies in elemental gadolinium,
\href{https://doi.org/10.1103/PhysRevB.105.104402}{Phys. Rev. B \textbf{105}, 104402 (2022)}.


\bibitem{parhizgar2013indirect}
F. Parhizgar, H. Rostami, and R. Asgari,
Indirect exchange interaction between magnetic adatoms in monolayer MoS$_2$,
\href{https://doi.org/10.1103/PhysRevB.87.125401}{Phys. Rev. B \textbf{87}, 125401 (2013)}.


\bibitem{geldenhuys1978rkky}
J. Geldenhuys and D. H. Wiid,
RKKY interaction and conduction electron polarisation,
\href{https://doi.org/10.1088/0305-4608/8/9/024}{J. Phys. F: Met. Phys. \textbf{8}, 2021--2033 (1978)}.


\bibitem{bisht2023investigation}
P. Bisht and R. N. Mahato,
Investigation of magnetic properties and colossal magnetoresistance in nanocrystalline doped manganite,
\href{https://doi.org/10.1088/1361-648X/acf01a}{J. Phys.: Condens. Matter \textbf{35}, 475802 (2023)}


\bibitem{chauhan2026magnetic}
H. C. Chauhan, U. C. Roy, S. Dan, A. Thamizhavel, and P. Das,
Magnetic measurements of Ni and Gd in the vicinity of the transition temperature,
\href{https://doi.org/10.1103/vp6c-6w9j}{Phys. Rev. B \textbf{113}, 024414 (2026)}.


\bibitem{Liu2017}
Y. Liu et al.,
Critical behavior of the van der Waals bonded ferromagnet
Fe$_{3-x}$GeTe$_2$,
\href{https://doi.org/10.1103/PhysRevB.96.144429}{Phys. Rev. B \textbf{96}, 144429 (2017)}.

\bibitem{liu2017critical}
B. Liu, Y. Zou, S. Zhou, L. Zhang, Z. Wang, H. Li, Z. Qu,
and Y. Zhang,
Critical behavior of the van der Waals bonded high-$T_C$
ferromagnet Fe$_3$GeTe$_2$,
\href{https://doi.org/10.1038/s41598-017-06671-5}{Sci. Rep. \textbf{7}, 6184 (2017)}.

\bibitem{ANPhysRevLett.19.786}
A. Arrott and J. E. Noakes,
Approximate equation of state for nickel near its critical temperature,
\href{https://doi.org/10.1103/PhysRevLett.19.786}{Phys. Rev. Lett. \textbf{19}, 786--789 (1967)}.


\bibitem{Kouvel1964}
J. S. Kouvel and M. E. Fisher,
Detailed magnetic behavior of nickel near its Curie point,
\href{https://doi.org/10.1103/PhysRev.136.A1626}{Phys. Rev. \textbf{136}, A1626--A1632 (1964)}.


\bibitem{Zhu2023}
L.-Z. Zhu, Q. Wang, S.-H. Zheng, P.-J. Wang, D. Ding, B.-Z. Tang,
P. Yu, J.-L. Yao, and L. Xia,
Effect of minor Ce substitution for Pr on the glass formability
and magnetocaloric effect of a Fe$_{88}$Zr$_4$Pr$_4$B$_4$
metallic glass,
\href{https://doi.org/10.3390/met13091531}{Metals \textbf{13}, 1531 (2023)}.


\bibitem{Gupta2018}
R. Gupta, I. N. Bhatti, and A. K. Pramanik,
Critical behavior in itinerant ferromagnet SrRu$_{1-x}$Ti$_x$O$_3$,
\href{https://doi.org/10.1016/j.jmmm.2018.05.105}{J. Magn. Magn. Mater. \textbf{465}, 193--200 (2018)}.


\bibitem{Tateiwa2014}
N. Tateiwa, Y. Haga, T. D. Matsuda, E. Yamamoto, and Z. Fisk,
Unconventional critical scaling of magnetization in ferromagnetic
uranium superconductors UGe$_2$ and URhGe,
\href{https://doi.org/10.1103/PhysRevB.89.064420}{Phys. Rev. B \textbf{89}, 064420 (2014)}.


\bibitem{Bhattacharyya2011}
A. Bhattacharyya, D. Jain, V. Ganesan, S. Giri, and S. Majumdar,
Investigation of weak itinerant ferromagnetism and critical behavior
of Y$_2$Ni$_7$,
\href{https://doi.org/10.1103/PhysRevB.84.184414}{Phys. Rev. B \textbf{84}, 184414 (2011)}.


\bibitem{yadav2026intrinsic}
N. Yadav, S. Kumawat, S. Soni, B. K. Mani, and P. Das,
Intrinsic intralayer antiferromagnetism in the near-stoichiometric
two-dimensional van der Waals magnet Fe$_3$GeTe$_2$,
\href{https://doi.org/10.1103/1fpk-fgkj}{Phys. Rev. B \textbf{114}, 154402 (2026)}.


\bibitem{deng2024magnetic}
R. Deng, B. Yu, H. Zheng, Y. Qiu, S. Wang, C. Ma, C. Wang,
Y. Zhu, and J. Fan,
Magnetic phase transition and critical behavior of the chiral magnet
$\beta$-Mn-type Co$_7$Zn$_7$Mn$_6$,
\href{https://doi.org/10.1016/j.cplett.2024.141622}{Chem. Phys. Lett. \textbf{856}, 141622 (2024)}.


\bibitem{zhang2013influence}
P. Zhang, P. Lampen, T. L. Phan, S. C. Yu, T. D. Thanh, N. H. Dan,
V. D. Lam, H. Srikanth, and M. H. Phan,
Influence of magnetic field on critical behavior near a first-order
transition in optimally doped manganites: The case of
La$_{1-x}$Ca$_x$MnO$_3$ ($0.2 \leq x \leq 0.4$),

\href{https://doi.org/10.1016/j.jmmm.2013.08.025}{J. Magn. Magn. Mater. \textbf{348}, 146--153 (2013)}.


\bibitem{wang2022competing}
Y. Wang, N. T. M. Duc, T. Feng, H. Wei, F. Qin, and M. H. Phan,
Competing ferromagnetic and antiferromagnetic interactions drive
the magnetocaloric tunability in Gd$_{55}$Co$_{30}$Ni$_x$Al$_{15-x}$
microwires,
\href{https://doi.org/10.1016/j.jallcom.2022.164328}{J. Alloys Compd. \textbf{907}, 164328 (2022)}.


\bibitem{clements2017critical}
E. M. Clements, R. Das, L. Li, P. J. Lampen-Kelley, M. H. Phan,
V. Keppens, D. Mandrus, and H. Srikanth,
Critical behavior and macroscopic phase diagram of the monoaxial
chiral helimagnet Cr$_{1/3}$NbS$_2$,
\href{https://doi.org/10.1038/s41598-017-06728-5}{Sci. Rep. \textbf{7}, 6545 (2017)}.


\bibitem{tozri2011critical}
A. Tozri, E. Dhahri, E. K. Hlil, and M. A. Valente,
Critical behavior near the paramagnetic to ferromagnetic phase
transition temperature in La$_{0.7}$Pb$_{0.05}$Na$_{0.25}$MnO$_3$,
\href{https://doi.org/10.1016/j.ssc.2010.11.036}{Solid State Commun. \textbf{151}, 315--320 (2011)}.


\bibitem{dash2019impression}
S. Dash, E. D. Baglasov, B. Weise, A. V. Lukoyanov, Y. V. Knyazev,
Y. I. Kuz'min, P. Kumar, and A. K. Patra,
Impression of magnetic clusters, critical behavior and magnetocaloric
effect in Fe$_3$Al alloys,
\href{https://doi.org/10.1039/C9CP01358J}{Phys. Chem. Chem. Phys. \textbf{21}, 10823--10833 (2019)}.


\bibitem{madhogaria2019evidence}
R. P. Madhogaria, R. Das, E. M. Clements, V. Kalappattil, M. H. Phan,
H. Srikanth, N. T. Dang, D. P. Kozlenko, and N. S. Bingham,
Evidence of long-range ferromagnetic order and spin frustration
effects in the double perovskite La$_2$CoMnO$_6$,
\href{https://doi.org/10.1103/PhysRevB.99.104436}{Phys. Rev. B \textbf{99}, 104436 (2019)}.


\bibitem{heilman2000field}
A. K. Heilman, Y. Y. Xue, B. Lorenz, M. Gospodinov, S. G. Dobreva,
and C. W. Chu,
Field-induced crossover in La$_{0.7}$Pb$_{0.3}$MnO$_3$,
\href{https://doi.org/10.1016/S0921-4534(00)00656-0}{Physica C \textbf{341--348}, 707--710 (2000)}.


\bibitem{wang2023field}
A. Wang, A. Rahman, Z. Du, J. Zhao, F. Meng, W. Liu, J. Fan,
C. Ma, M. Ge, L. Pi, L. Zhang, and Y. Zhang,
Field-dependent anisotropic room-temperature ferromagnetism in
Cr$_3$Te$_4$,
\href{https://doi.org/10.1103/PhysRevB.108.094429}{Phys. Rev. B \textbf{108}, 094429 (2023)}.


\bibitem{chauhan2022different}
H. C. Chauhan, B. Kumar, A. Tiwari, J. K. Tiwari, and S. Ghosh,
Different Critical Exponents on Two Sides of a Transition: Observation of Crossover from Ising to Heisenberg Exchange in Skyrmion Host Cu$_2$OSeO$_3$,
\href{https://doi.org/10.1103/PhysRevLett.128.015703}{Phys. Rev. Lett. \textbf{128}, 015703 (2022)}.


\bibitem{yi2016competing}
J. Yi, H. Zhuang, Q. Zou, Z. Wu, G. Cao, S. Tang, S. A. Calder,
P. R. C. Kent, D. Mandrus, and Z. Gai,
Competing antiferromagnetism in a quasi-2D itinerant ferromagnet:
Fe$_3$GeTe$_2$,
\href{https://doi.org/10.1088/2053-1583/4/1/011005}{2D Mater. \textbf{4}, 011005 (2016)}.


\bibitem{Deng2018}
Y. Deng, Y. Yu, Y. Song, J. Zhang, N. Z. Wang, Z. Sun, Y. Yi,
Y. Z. Wu, S. Wu, J. Zhu, J. Wang, X. H. Chen, and Y. Zhang,
Gate-tunable room-temperature ferromagnetism in two-dimensional
Fe$_3$GeTe$_2$,
\href{https://doi.org/10.1038/s41586-018-0626-9}{Nature \textbf{563}, 94--99 (2018)}.


\bibitem{Tan2018}
C. Tan, J. Lee, S.-G. Jung, T. Park, S. Albarakati, J. Partridge,
M. R. Field, D. G. McCulloch, L. Wang, and C. Lee,
Hard magnetic properties in nanoflake van der Waals Fe$_3$GeTe$_2$,
\href{https://doi.org/10.1038/s41467-018-04018-w}{Nat. Commun. \textbf{9}, 1554 (2018)}.


\bibitem{Fei2018}
Z. Fei, B. Huang, P. Malinowski, W. Wang, T. Song, J. Sanchez,
W. Yao, D. Xiao, X. Zhu, A. F. May, W. Wu, D. H. Cobden,
J.-H. Chu, and X. Xu,
Two-dimensional itinerant ferromagnetism in atomically thin
Fe$_3$GeTe$_2$,
\href{https://doi.org/10.1038/s41563-018-0149-7}{Nat. Mater. \textbf{17}, 778--782 (2018)}.


\bibitem{10.1063/1.4961592}
N. León-Brito, E. D. Bauer, F. Ronning, J. D. Thompson,
and R. Movshovich,
Magnetic microstructure and magnetic properties of uniaxial
itinerant ferromagnet Fe$_3$GeTe$_2$,
\href{https://doi.org/10.1063/1.4961592}{J. Appl. Phys. \textbf{120}, 083903 (2016)}.

\bibitem{AnkitaPhysRevB.109.L020407}
A. Tiwari, H. Ahn, B. Kumar, J. Saini, P. K. Srivastava, B. Singh,
C. Lee, and S. Ghosh,
Phase transition in the two-dimensional Heisenberg ferromagnet
Fe$_3$GeTe$_2$ with long-range interaction,
\href{https://doi.org/10.1103/PhysRevB.109.L020407}{Phys. Rev. B \textbf{109}, L020407 (2024)}.

\bibitem{Kim2026}
H. Kim, K.-X. Zhang, Y.-H. Li, G. Park, R. Cheng, and J.-G. Park,
Emergent giant topological Hall effect in twisted Fe$_3$GeTe$_2$
metallic system,
\href{https://doi.org/10.1038/s41467-026-69454-5}{Nat. Commun. \textbf{17}, 2931 (2026)}.


\bibitem{zhang2018emergence}
Y. Zhang, H. Lu, X. Zhu, S. Tan, W. Feng, Q. Liu, W. Zhang,
Q. Chen, Y. Liu, et al.,
Emergence of Kondo lattice behavior in a van der Waals itinerant
ferromagnet, Fe$_3$GeTe$_2$,
\href{https://doi.org/10.1126/sciadv.aao6791}{Sci. Adv. \textbf{4}, eaao6791 (2018)}.


\bibitem{sharma2023non}
P. R. Sharma, T. W. Kim, and H. Noh,
Non-monotonic temperature dependence of anomalous Hall effect
in a 2D van der Waals ferromagnetic material Fe$_3$GeTe$_2$,
\href{https://doi.org/10.1016/j.matchemphys.2023.127738}{Mater. Chem. Phys. \textbf{302}, 127738 (2023)}.


\bibitem{PhysRevLett.17.1133}
N. D. Mermin and H. Wagner,
Absence of Ferromagnetism or Antiferromagnetism in One- or Two-Dimensional Isotropic Heisenberg Models,
\href{https://doi.org/10.1103/PhysRevLett.17.1133}{Phys. Rev. Lett. \textbf{17}, 1133--1136 (1966)}.


\bibitem{PhysRev.158.383}
P. C. Hohenberg,
Existence of Long-Range Order in One and Two Dimensions,
\href{https://doi.org/10.1103/PhysRev.158.383}{Phys. Rev. \textbf{158}, 383--386 (1967)}.


\bibitem{Law2018}
J. Y. Law, V. Franco, L. M. Moreno-Ram{\'i}rez, A. Conde,
D. Y. Karpenkov, I. Radulov, K. P. Skokov, and O. Gutfleisch,
A quantitative criterion for determining the order of magnetic
phase transitions using the magnetocaloric effect,
\href{https://doi.org/10.1038/s41467-018-05111-w}{Nat. Commun. \textbf{9}, 2680 (2018)}.


\bibitem{Franco2006}
V. Franco, J. S. Bl{\'a}zquez, and A. Conde,
Field dependence of the magnetocaloric effect in materials with
a second-order phase transition: A master curve for the magnetic
entropy change,
\href{https://doi.org/10.1063/1.2399361}{Appl. Phys. Lett. \textbf{89}, 222512 (2006)}.


\bibitem{toledano1987landau}
J. C. Tol{\'e}dano and P. Tol{\'e}dano,
The Landau Theory of Phase Transitions: Application to Structural, Incommensurate, Magnetic and Liquid Crystal Systems,
World Scientific (1987).

\bibitem{SM} See Supplemental Material at www.yyy.com for discussion on the intercept of the MAPs and normalized slope plots, the modified Arrott plots and their derivatives.

\bibitem{GarciaGonzalo2003}
J. Garc{\'i}a and J. A. Gonzalo,
Accurate Monte Carlo critical exponents for Ising lattices,
\href{https://doi.org/10.1016/S0378-4371(03)00362-5}{Physica A \textbf{326}, 464--472 (2003)}.


\bibitem{GarciaMarquesGonzalo2012}
J. G. Garc{\'i}a, M. I. Marqu{\'e}s, and J. A. Gonzalo,
Monte Carlo study of Ising critical isotherms,
\href{https://doi.org/10.1080/00150193.2012.671731}{Ferroelectrics \textbf{426}, 166--170 (2012)}.


\bibitem{PhysRevLett.29.917}
M. E. Fisher, S.-k. Ma, and B. G. Nickel,
Critical Exponents for Long-Range Interactions,
\href{https://doi.org/10.1103/PhysRevLett.29.917}{Phys. Rev. Lett. \textbf{29}, 917--920 (1972)}.


\bibitem{TIWARI2021168020}
J. K. Tiwari, B. Kumar, H. C. Chauhan, and S. Ghosh,
Critical scaling and magnetic phase diagram of bi-skyrmion-hosting
quasi-two-dimensional La$_{1.37}$Sr$_{1.63}$Mn$_2$O$_7$
bi-layer manganite,
\href{https://doi.org/10.1016/j.jmmm.2021.168020}{J. Magn. Magn. Mater. \textbf{535}, 168020 (2021)}


\bibitem{BHATTACHARYA2023170274}
A. Bhattacharya, A. Ahmed, S. DuttaGupta, and I. Das,
Critical behavior and phase diagram of skyrmion-hosting material
Co$_{3.6}$Fe$_{4.4}$Zn$_8$Mn$_4$ probed by anomalous Hall effect,
\href{https://doi.org/10.1016/j.jallcom.2023.170274}{J. Alloys Compd. \textbf{960}, 170274 (2023)}.


\bibitem{PhysRevB.65.214424}
D. Kim, B. L. Zink, F. Hellman, and J. M. D. Coey,
Critical behavior of La$_{0.75}$Sr$_{0.25}$MnO$_3$,
\href{https://doi.org/10.1103/PhysRevB.65.214424}{Phys. Rev. B \textbf{65}, 214424 (2002)}.


\bibitem{KAUL19855}
S. N. Kaul,
Static critical phenomena in ferromagnets with quenched disorder,
\href{https://doi.org/10.1016/0304-8853(85)90128-3}{J. Magn. Magn. Mater. \textbf{53}, 5--53 (1985)}.


\bibitem{PhysRevB.21.3976}
J. C. Le Guillou and J. Zinn-Justin,
Critical exponents from field theory,
\href{https://doi.org/10.1103/PhysRevB.21.3976}{Phys. Rev. B \textbf{21}, 3976--3998 (1980)}.


\bibitem{PhysRevLett.115.200601}
F. L{\'e}onard and B. Delamotte,
Critical Exponents Can Be Different on the Two Sides of a Transition: A Generic Mechanism,
\href{https://doi.org/10.1103/PhysRevLett.115.200601}{Phys. Rev. Lett. \textbf{115}, 200601 (2015)}.


\end{thebibliography}
\end{document}